\documentclass[conference]{IEEEtran}
\IEEEoverridecommandlockouts
\usepackage{cite}
\usepackage{amsmath,amssymb,amsfonts}
\usepackage{amsmath}

\usepackage{graphicx}
\usepackage{textcomp}
\usepackage{xcolor}
\usepackage{graphicx}
\usepackage{float} 
\usepackage{subfigure}
\usepackage{amsmath}
\usepackage{amsfonts,amssymb}
\usepackage{mathrsfs}
\usepackage{mathtools}
\usepackage{algorithm}
\usepackage{algorithmicx}
\usepackage{algpseudocode}
\usepackage{bm}
\bibliography{IEEEabrv}
\def\BibTeX{{\rm B\kern-.05em{\sc i\kern-.025em b}\kern-.08em
    T\kern-.1667em\lower.7ex\hbox{E}\kern-.125emX}}

\begin{document}

\title{Ergodic H-S/MRC Mutual Information\\
}

\author{\IEEEauthorblockN{Zeliang Ou, Chongjun Ouyang, Pei Yang, Lu Zhang and Hongwen Yang}
\IEEEauthorblockA{\textit{Wireless Theories and Technologies Lab} \\
\textit{Beijing University of Posts and Telecommunications}\\
Beijing, China \\
\{ouzeliang, DragonAim, yp, zhangl\_96, yanghong\}@bupt.edu.cn}
}

\maketitle

\begin{abstract}
This paper studies the ergodic mutual information of hybrid selection/maximal-ratio combining (H-S/MRC) diversity system under BPSK/QPSK modulations. We consider a simple single-input multiple-output (SIMO) channel, where a subset of branches are selected and combined using maximal-ratio combining (MRC) to maximize the instantaneous Signal to Noise Ratio (SNR) at the receiver. For independent and identically distributed (i.i.d.) Rayleigh flat fading, a general recursive expression is developed to estimate the ergodic input-output mutual information of the whole system. Besides analytical derivations, simulations are provided to demonstrate the feasibility and validity of the derived results.
\end{abstract}


\section{Introduction}

The concept of hybrid selection/maximal-ratio combining (H-S/MRC) diversity system was first proposed by Moe Z. Win and Jack H. Winters in 1999 \cite{b1}. In this system, a subset of antennas with the largest Signal to Noise Ratio (SNR) are selected to receive messages at each instant, and these branches are combined using MRC to maximize the instantaneous SNR. Under single-input multiple-output (SIMO) channels, this H-S/MRC technology is usually adopted to alleviate the requirement on radio-frequency (RF) chains but keep considerably high spectral efficiency.

Since 1999, many researches on the H-S/MRC system were presented\cite{b2,b3,b4,b5,b6,b7,b8,b9}. For example, Win {\emph{et al.}}, in \cite{b2} derived the mean and variance of the received SNR in H-S/MRC systems based on the novel virtual branch method. Later, this method was further used to formulate closed-from expressions of symbol error rate (SER) for H-S/MRC system under Rayleigh fading \cite{b3}. After that, more formulas of SER were continuously presented for different multi-path fading types  \cite{b4,b5,b6}. Besides the rich results on SER, theoretical results on the input-output mutual information (MI) of H-S/MRC system were also widely studied in the literature\cite{b7,b8,b9}. Molisch {\emph{et al.}}, in \cite{b7}, proposed an analytical expression of the ergodic MI over Nakagami-$m$ fading. Furthermore, the works in \cite{b8} and \cite{b9} investigated the asymptotic ergodic MI when the receiver was equipped with large-scale antenna array. Nevertheless, nearly all these aforementioned works have assumed that the input signals followed Gaussian distribution. Actually, a significant situation which must be studied when considering a practical communication system is the scenario when the channel inputs are drawn from discrete constellations. 

However, existing literature on the ergodic MI, under finite-alphabet inputs, failed to formulate any closed-form expressions for it. Motivated by this, this paper concentrates on the ergodic mutual information under Rayleigh fading and formulates a recursive expression for the ergodic mutual information on the basis of the important derivation of H-S/MRC system, ever developed in \cite{b3}, and the recent development in ergodic mutual information of single-input single-output (SISO) channels under Nakagami-$m$ fading. To the best of our knowledge, this is the first time to propose a recursive expression of ergodic MI for  H-S/MRC systems under finite-alphabet inputs. For the sake of brevity, assume that the modulation mode is BPSK/QPSK. 

The remaining parts of this manuscript are structured as follows: Section \ref{section2} describes the system model. In Section \ref{section3}, the recursive expression for the ergodic mutual information is derived. Then, two special cases are analyzed in Section \ref{section4}. Finally, Section \ref{section5} concludes the paper.

\section{System Model}
\label{section2}
Consider a SIMO system, where the transmitter and receiver are respectively equipped with 1 and $N_{\rm{r}}$ antennas. At the receiver, the received signal vector is given by
\begin{equation}
\label{eq2_1}
{{\bf{y}}}={{\bf{h}}}{x}+{{\bf{w}}},
\end{equation}
where ${{x}}$ is the transmitted signal constrained by finite constellation size, such as BPSK, and unit power; ${{\bf{w}}}{\sim}{\mathcal{CN}}{({\bf{0}},{{\bf{I}}_{N_{\rm{r}}}})}$ is the additive white Gaussian noise (AWGN). Suppose that the channel suffers from independent and identically distributed (i.i.d.) Rayleigh flat fading, in other words, the magnitude of each element in the channel matrix ${\bf{h}}{\in}{{\mathbb{C}}^{{N_{\rm{r}}}{\times}{1}}}$ follows i.i.d. Rayleigh distribution with the same probability density function (PDF) as follows:
\begin{equation}
\label{eq2_2}
f(x)=\frac{2x}{\bar\gamma}{\rm{e}}^{\frac{-x^2}{\bar\gamma}},
\end{equation}
where $\bar{\gamma}$ is the average per-antenna or per-branch SNR at the receiver.

Now, assume that $L$ ($L\leq N_{\rm{r}}$) out of $N_{\rm{r}}$ receive antennas corresponding to the strongest $L$ branches are activated. Additionally, assume that the channel state information (CSI) is only available at the receiver and maximal-ratio combining is applied. Let $\widetilde{\bf{h}}$, $\widetilde{\bf{y}}$ and $\widetilde{\bf{w}}$ denote the channel matrix, the received signal and the additive noise after antenna selection, thus $\widetilde{\bf{y}}=\widetilde{\bf{h}}x+\widetilde{\bf{w}}$ holds. Moreover, the result obtained from MRC can be written as\cite{b10}
\begin{equation}
\label{eq2_3}
{\widehat{\rm{y}}}=\frac{\widetilde{\bf{h}}^{\dagger}}{||\widetilde{\bf{h}}||}\widetilde{\bf{y}}={||\widetilde{\bf{h}}||}x+{\widehat{\rm{w}}},
\end{equation}
in which ${\widehat{\rm{w}}}=\frac{\widetilde{\bf{h}}^{\dagger}}{||\widetilde{\bf{h}}||}\widetilde{\bf{w}}\sim{\mathcal{CN}}{({0,1})}$ and $(\cdot)^{\dagger}$ is the Hermitian operator.

As stated before, the input signals, modulated by BPSK or QPSK, follow non-Gaussian distribution, thus the classical Shannon formula $\log_2\left(1+{\rm{SNR}}\right)$ can not be directly utilized to measure the input-output mutual information of this H-S/MRC system. Furthermore, it is sufficient to consider BPSK since QPSK is the superposition of two orthogonal BPSK modulations. Consequently, we assume that the transmitted data stream are i.i.d. zero-mean binary symbols with equal probabilities, and the input-output MI (in ``nats/symbol'') in terms of the SNR $\gamma$ under BPSK modulation over AWGN channels can be formulated as \cite{b11}
\begin{equation}
\label{Equ4}
{\mathcal{I}}\left(\gamma\right)=\ln2-\int_{-\infty}^{+\infty}{\frac{1}{\sqrt{2\pi}}{\rm{e}}^{-\frac{u^2}{2}}\ln\left(1+{\rm{e}}^{-2\sqrt{\gamma}u-2\gamma}\right){\rm{d}}u}.
\end{equation} 
Additionally, the equation ${{I}}\left(x;{\widehat{\rm{y}}}\right)={{I}}\left(x;\widetilde{\bf{{y}}}\right)$ holds, for the MRC is a lossless operation i.e., $\widetilde{\bf{y}}$ can be generated from ${\widehat{\rm{y}}}$. Define $\Gamma=||\widetilde{\bf{h}}||^2$ as the received SNR after antenna selection, the ergodic input-output mutual information of BPSK in the H-S/MRC system can be written as
\begin{equation}
\label{Equ5}
\begin{split}
\bar{C}=&{\mathbb{E}}\left[I\left(x;\widetilde{\bf{{y}}}\right)\right]={\mathbb{E}}\left[{\mathcal{I}}{\left(||\widetilde{\bf{h}}||^2\right)}\right]=\int_{0}^{+\infty}{{{\mathcal{I}}(\gamma)}F_{\Gamma}(\gamma){\rm{d}}\gamma}\\
=&\ln{2}-\int\limits_{0}^{+\infty}\int\limits_{-\infty}^{+\infty}{\frac{{\rm{e}}^{-\frac{u^2}{2}}}{\sqrt{2\pi}}\ln\left(1+{\rm{e}}^{-2\sqrt{\gamma}u-2\gamma}\right)}F_{\Gamma}(\gamma){\rm{d}}u{\rm{d}}\gamma,
\end{split}
\end{equation}
where $F_{\Gamma}\left(\cdot\right)$ denotes the PDF of $\Gamma$. 

\section{Ergodic Mutual Information}
\label{section3}
In the following section, we will first investigate the expression for $F_{\Gamma}\left(\cdot\right)$ and then derive the analytical expression for $\bar{C}$. 
According to the derivation in \cite{b3}, the characteristic function of $F_{\Gamma}\left(\cdot\right)$ can be written as
\begin{equation}
\label{eq2_1}
\begin{split}
\Psi_{\Gamma}(\phi)=\left(\frac{1}{1-\phi\bar\gamma}\right)^{L}\prod_{n=L+1}^{N_{\rm{r}}}\left(\frac{1}{1-\phi\bar\gamma\frac{L}{n}}\right).\\
\end{split}
\end{equation}
Moreover, Equ. \eqref{eq2_1} can be rewritten into the simple canonical structure. On the basis of partial fraction decomposition (PFD), the canonical expansion of 
\begin{equation}
\label{eq2_2}
t(x)=\prod_{n=1}^{N}\left(\frac{c_{n}}{c_{n}+x}\right)^{\mu_{n}}
\end{equation}
where the $\{-c_{n}\}$ are the $N$ poles of $t(x)$, each having algebraic multiplicity $\mu_n$, 
can be expressed as
\begin{equation}
\label{eq2_3}
t(x)=\sum_{n=1}^{N}\sum_{k=1}^{\mu_n}A_{n,k}\left(\frac{c_{n}}{c_{n}+x}\right)^{k}.
\end{equation} 
And the weighting coefficients of the canonical expansion are given by
\begin{equation}
\label{eq2_4}
\begin{split}
A_{n,k}=&\frac{1}{c_n^k\left(\mu_n-k\right)!}t_n^{\left(\mu_n-k\right)}\left(0\right),\\
&\qquad n=1,\cdots,N,\quad k=1,\cdots,\mu_n
\end{split}
\end{equation} 
where $t_{n}^{\left(k\right)}\left(0\right)$ represents the $k$th derivation of $t_{n}{\left(x\right)}=x^{\mu_n}t\left(x-c_n\right)$ evaluated at $x=0$. According to Equ. \eqref{eq2_2} and Equ. \eqref{eq2_3}, the canonical structure of Equ. \eqref{eq2_1} can be expressed as
\begin{equation}
\label{eq2_5}
\begin{split}
\Psi&_{\Gamma}(j\omega)=\\
&\sum_{k=1}^{L}A_{1,k}\left(\frac{1}{1-j\omega\bar\gamma}\right)^{k}+\sum_{n=2}^{N_{\rm{r}}-L+1}A_{n,1}\left(\frac{1}{1-j\omega\bar\gamma\frac{L}{L+n-1}}\right).
\end{split}
\end{equation}
Under this circumstance, the coefficients $\{c_{\rm{n}}\}$ and $\{\mu_{\rm{n}}\}$ in $A_{n,k}$ are given by
\begin{equation}
\label{eq2_6}
\begin{split}
&\mu_n=
\begin{cases}
L& {n=1}\\
1& {n=2,\cdots,N_{\rm{r}}-L+1}
\end{cases}
\\
&c_n=
\begin{cases}
{-1}/{\bar\gamma}& {n=1}\\
\frac{-\left(L+n-1\right)}{L\bar\gamma}& {n=2,\cdots,N_{\rm{r}}-L+1}.
\end{cases}
\end{split}
\end{equation}
To obtain $F_{\Gamma}\left(\cdot\right)$, it is necessary to apply Fourier transform into $\Psi_{\Gamma}(j\omega)$, thus the PDF of the received SNR can be written as
\begin{equation}
\label{eq2_5}
\begin{split}
F_{\Gamma}\left(\gamma\right)=\sum_{k=1}^{L}&A_{1,k}\frac{\gamma^{k-1}}{\Gamma\left(k\right)\bar{\gamma}^k}{\rm{e}}^{-\frac{\gamma}{\bar{\gamma}}}\\
&+\sum_{n=2}^{N_{\rm{r}}-L+1}A_{n,1}\frac{L+n-1}{\bar\gamma L}{\rm{e}}^{-\frac{\gamma}{\bar\gamma L/\left(L+n-1\right)}}.
\end{split}
\end{equation}
As a result, the ergodic mutual information can be summarized as
\begin{equation}
\label{eq13}
\begin{split}
\bar{C}=&\sum_{k=1}^{L}A_{1,k}\underbrace{\int_{0}^{+\infty}{\mathcal{I}}\left(\gamma\right)\frac{\gamma^{k-1}}{\Gamma\left(k\right)\bar{\gamma}^k}{\rm{e}}^{-\frac{\gamma}{\bar{\gamma}}}{\rm{d}}\gamma}_{\Lambda_1}\\
&+\sum_{n=2}^{N_{\rm{r}}-L+1}A_{n,1}\underbrace{\int_{0}^{+\infty}{\mathcal{I}}\left(\gamma\right)\frac{L+n-1}{\bar\gamma L}{\rm{e}}^{-\frac{\gamma\left(L+n-1\right)}{\bar\gamma L}}{\rm{d}}\gamma}_{\Lambda_2}.
\end{split}
\end{equation}
Notably, in Equ. \eqref{eq13}, $\Lambda_1$ and $\Lambda_2$ can be treated as the ergodic mutual information over SISO channels under Nakagami-$m$ and Rayleigh fading, respectively. Meanwhile, Rayleigh fading can be regarded as Nakagami-1 fading. Consequently, the ergodic mutual information of H-S/MRC systems under Rayleigh fading can be equivalent to the superposition of the mutual information of SISO systems under Nakagami-$m$ fading. To this end, it is necessary to derive the ergodic mutual information of SISO systems under Nakagami-$m$ fading. 
However, there is no closed-form solution to the ergodic mutual information in Nakagami-$m$ fading due to the complicated form of Equ. \eqref{Equ4}. 

Fortunately, a recursive expression for the ergodic Nakagami-$m$ SISO mutual information can be obtained even though the closed-form expression is hard to figure out, that is\cite{b12}
\begin{equation}
\label{eq4_2}
{\mathcal{I}}_{\rm{B}}\left(m,\bar\gamma\right)=\ln{2}-{\mathcal{F}}(m,m,\bar\gamma),  
\end{equation}
in which $\bar\gamma$ denotes the average per-antenna SNR and 
\begin{equation}
\label{eq4_3}
\begin{split}
{\mathcal{F}}(m,n,\bar\gamma)=&\frac{2{\mathcal{H}}_1m^m}{\bar\gamma^m\Gamma(m)}+\frac{\left(n-1\right)\bar\gamma}{m}{\mathcal{F}}(m,n-1,\bar\gamma)\\
&-\frac{\bar\gamma}{2m}\sum_{k=0}^{\infty}(-1)^k{\mathcal{B}}_k(m,n,\bar\gamma),  
\end{split}
\end{equation}
where ${\mathcal{B}}_k(m,n,\bar\gamma)$ satisfies the following relationship 
\begin{equation}
\label{eq4_4}
\begin{split}
&{\mathcal{B}}_k(m,n,\bar\gamma)\left(1-\frac{(2k+1)^2\bar\gamma}{2(m+0.5\bar\gamma)}\right)\\=&\frac{{\mathcal{H}}_2m^m}{\bar\gamma^m\Gamma(m)}+\frac{\left(n-1\right)\bar\gamma}{m+0.5\bar\gamma}{\mathcal{B}}_k(m,n-1,\bar\gamma)\\
&-\frac{\sqrt{2}(2k+1)m^m(2n-3)!!}{2^{n-1}\bar\gamma^{m-n-0.5}\Gamma(m)(m+0.5\bar\gamma)^{n+0.5}},  
\end{split}
\end{equation}
and
\begin{equation}
\label{eq4_5}
{\mathcal{H}}_1=\begin{cases}
\frac{\bar\gamma\ln{2}}{2m},&n=1 \\
0& n\neq1
\end{cases},
\end{equation}
\begin{equation}
\label{eq4_6}
{\mathcal{H}}_2=\begin{cases}
\frac{2\bar\gamma}{m+0.5\bar\gamma},&n=1 \\
0& n\neq1
\end{cases}.
\end{equation}
Besides, ${\mathcal{F}}\left(m,0,\bar\gamma\right)={\mathcal{B}}_{k}\left(m,0,\bar\gamma\right)=0$ for $k=0$.
Therefore, ergodic H-S/MRC mutual information can be also represented into a recursive form as follows:
\begin{equation}
\bar{C}=\sum_{k=1}^{L}A_{1,k}{\mathcal{I}}_{\rm{B}}\left(k,k\bar\gamma\right)+\sum_{n=2}^{N_{\rm{r}}-L+1}A_{n,1}{\mathcal{I}}_{\rm{B}}\left(1,\frac{\bar\gamma L}{L+n-1}\right).
\end{equation}

\section{Special Cases}
\label{section4}
\subsection{$L=1$}
When $L=1$, the H-S/MRC system degrades into the system with selection combination (SC). Under this circumstance, the ergodic mutual information is formulated as:
\begin{equation}
\bar{C}=\sum_{n=1}^{N_{\rm{r}}}A_{n,1}{\mathcal{I}}_{\rm{B}}\left(1,\bar\gamma /n\right)=\sum_{n=1}^{N_{\rm{r}}}A_{n,1}\left[\ln{2}-{\mathcal{F}}\left(1,1,\bar\gamma /n\right)\right].
\end{equation} 
On the basis of Equ. \eqref{eq2_4}, the coefficients are derived as
\begin{equation}\
\begin{split}
A_{n,1}=&\left(\prod_{i=1}^{n-1}\frac{c_i}{c_i-c_n}\right)\left(\prod_{i=n+1}^{N_{\rm{r}}}\frac{c_i}{c_i-c_n}\right)\\
=&\left(\prod_{i=1}^{n-1}\frac{i}{i-n}\right)\left(\prod_{i=n+1}^{N_{\rm{r}}}\frac{i}{i-n}\right).
\end{split}
\end{equation}
Moreover, by Equ. \eqref{eq4_3}, we have ${\mathcal{F}}\left(1,1,\bar\gamma\right)=\ln{2}-\frac{\bar\gamma}{2}\sum_{k=0}^{\infty}(-1)^k{\mathcal{B}}_k(1,1,\bar\gamma)$. Next, let us turn to ${\mathcal{B}}_k(1,1,\bar\gamma)$. On the basis of Equ. \eqref{eq4_4}, ${\mathcal{B}}_k\left(m,1,\bar\gamma\right)$ can be simplified as follows:
\begin{equation}
{\mathcal{B}}_k\left(m,1,\bar\gamma\right)=\sqrt{\frac{4\bar\gamma}{2m+\bar\gamma}}\frac{m^m}{\bar\gamma^m\Gamma\left(m\right)}\frac{1}{k+T_m\left(\bar\gamma\right)},
\end{equation}
and
\begin{equation}
T_m\left(\bar\gamma\right)=\frac{1}{2}\left(1+\sqrt{1+\frac{2m}{\bar\gamma}}\right).
\end{equation}
Therefore, ${\mathcal{B}}_k\left(\cdot,\cdot,\cdot\right)$ terms, when $m=1$ and $n=1$, are given by
\begin{equation}
{\mathcal{B}}_k(1,1,\bar\gamma)=\sqrt{\frac{4\bar\gamma}{2+\bar\gamma}}\frac{1}{\bar\gamma}\frac{1}{k+T_1\left(\bar\gamma\right)}.
\end{equation}
In summary, the final result for the ergodic mutual information of BPSK can be derived as
\begin{equation}
\label{eq25}
\bar{C}=\sum_{n=1}^{N_{\rm{r}}}A_{n,1}\sqrt{\frac{\bar\gamma/n}{2+\bar\gamma/n}}\left[\sum_{k=0}^{+\infty}\frac{\left(-1\right)^k}{k+T_1\left(\bar\gamma/n\right)}\right].
\end{equation} 
To accelerate the convergence of the series in Equ. \eqref{eq25}, the following formula of $\beta-$function can be utilized, which satisfies the following expression \cite[Equ. (8.372)]{b13}:
\begin{equation}
\label{eq26}
\begin{split}
\beta(x)\overset{(1)}{=}&\sum_{k=0}^{+\infty}\frac{(-1)^k}{x+k}\overset{(2)}{=}\sum_{k=0}^{+\infty}\frac{2^{-(k+1)}k!}{x(x+1)(x+2)\cdots(x+k)}.
\end{split}
\end{equation}  
In Equ. \eqref{eq26}, (1) is the definition formula and (2) is the expansion.
Due to the term $2^{-\left(k+1\right)}$, the expansion will decreases exponentially, causing the fast convergence speed in the final result.
By Equ. \eqref{eq26}, Equ. \eqref{eq25} becomes
\begin{equation}
\label{eq27}
\begin{split}
\bar{C}=&\sum_{n=1}^{N_{\rm{r}}}A_{n,1}\sqrt{\frac{\bar\gamma/n}{2+\bar\gamma/n}}\\
&\times\left[\sum_{k=0}^{+\infty}\frac{2^{-(k+1)}k!}{T_1\left(\frac{\bar\gamma}{n}\right)(T_1\left(\frac{\bar\gamma}{n}\right)+1)\cdots(T_1\left(\frac{\bar\gamma}{n}\right)+k)}\right]
\end{split}
\end{equation}

\subsection{$L=2$}
For $L=2$, the ergodic H-S/MRC mutual information can be formulated as
\begin{equation}
\label{eq28}
\begin{split}
\bar{C}=&A_{1,2}{\mathcal{I}}_{\rm{B}}\left(2,2\bar\gamma\right)+\sum_{n=1}^{N_{\rm{r}}-1}A_{n,1}{\mathcal{I}}_{\rm{B}}\left(1,2\bar\gamma /(n+1)\right)\\
=&A_{1,2}\left[\ln{2}-{\mathcal{F}}\left(2,2,2\bar\gamma\right)\right]\\
&+\sum_{n=1}^{N_{\rm{r}}-1}A_{n,1}\left[\ln{2}-{\mathcal{F}}\left(1,1,\frac{2\bar\gamma}{n+1}\right)\right].
\end{split}
\end{equation}
On the basis of Equ. \eqref{eq2_4}, the coefficients are derived as
\begin{subequations}
\begin{align}
A_{1,2}&=\prod_{i=2}^{N_{\rm{r}}-1}\left(\frac{i+1}{i-1}\right),\\
A_{1,1}&=-{N_{\rm{r}}}\left({N_{\rm{r}}-1}\right)\sum_{i=2}^{N_{\rm{r}}-1}\frac{1}{i-1},\\
A_{n,1}&=\left(\frac{2}{1-n}\right)^2\left(\prod_{i=2}^{n-1}\frac{i+1}{i-n}\right)\left(\prod_{i=n+1}^{N_{\rm{r}}-1}\frac{i+1}{i-n}\right)\quad\left(n>1\right).
\end{align}
\end{subequations}
Since ${\mathcal{F}}\left(1,1,\bar\gamma\right)$ has been solved in the last subsection, it is sufficient to calculate ${\mathcal{F}}\left(2,2,\bar\gamma\right)$.
\begin{subequations}
\label{eq30}
\begin{align}
{\mathcal{F}}\left(2,1,\bar\gamma\right)&=\frac{2\ln{2}}{\bar\gamma}-\frac{\bar\gamma}{4}\sum_{k=0}^{+\infty}{\mathcal{B}}_k\left(2,1,\bar\gamma\right),\\
{\mathcal{F}}\left(2,2,\bar\gamma\right)&=\frac{\bar\gamma}{2}{\mathcal{F}}\left(2,1,\bar\gamma\right)-\frac{\bar\gamma}{4}\sum_{k=0}^{+\infty}{\mathcal{B}}_k\left(2,2,\bar\gamma\right).
\end{align}
\end{subequations}
Furthermore,
\begin{subequations}
\label{eq31}
\begin{align}
{\mathcal{B}}_k\left(2,1,\bar\gamma\right)&=\frac{8}{\bar\gamma^2}\sqrt{\frac{4\bar\gamma}{4+\bar\gamma}}\frac{1}{k+T_2\left(\bar\gamma\right)},\\
{\mathcal{B}}_k\left(2,2,\bar\gamma\right)&=\frac{8\bar\gamma^{-0.5}}{\left(4+\bar\gamma\right)^{1.5}}\frac{1}{k+T_2\left(\bar\gamma\right)}+\frac{4\bar\gamma^{-1}}{4+\bar\gamma}\frac{1}{\left(k+T_2\left(\bar\gamma\right)\right)^2}.
\end{align}
\end{subequations}
Substituting Equ. \eqref{eq30} and Equ. \eqref{eq31} into Equ. \eqref{eq28}, the expression for the ergodic mutual information can be developed after some basic mathematical manipulations, that is
\begin{equation}
\begin{split}
\bar{C}&=A_{1,2}\left(\frac{1}{4+2\bar\gamma}\sum_{k=0}^{+\infty}\frac{\left(-1\right)^k}{\left[k+T_2\left(2\bar\gamma\right)\right]^2}\right)\\
&+A_{1,2}\left(\sqrt{\frac{\bar\gamma}{2+\bar\gamma}}\frac{3+\bar\gamma}{2+\bar\gamma}\sum_{k=0}^{+\infty}\frac{\left(-1\right)^k}{k+T_2\left(2\bar\gamma\right)}\right)+\\
&\sum_{n=1}^{N_{\rm{r}}-1}A_{n,1}\sqrt{\frac{2\bar\gamma /(n+1)}{2+2\bar\gamma /(n+1)}}\left[\sum_{k=0}^{+\infty}\frac{\left(-1\right)^k}{k+T_1\left(2\bar\gamma /(n+1)\right)}\right].
\end{split}
\end{equation}
Similar to Equ. \eqref{eq27}, to accelerate the convergence, the $\beta$-function can be continuously used, the final result is shown in Equ. \eqref{eq33}.
\begin{figure*}
\setlength{\abovecaptionskip}{-4pt} 
\begin{equation}
\label{eq33}
\begin{split}
\bar{C}=&A_{1,2}\left(\frac{1}{4+2\bar\gamma}\underbrace{\sum_{k=0}^{+\infty}\frac{2^{-\left(k+1\right)}k!}{T_2\left(2\bar\gamma\right)\cdots\left[T_2\left(2\bar\gamma\right)+k\right]}\sum_{i=0}^{k}\frac{1}{T_2\left(2\bar\gamma\right)+i}}_{\Xi_1}+\sqrt{\frac{\bar\gamma}{2+\bar\gamma}}\frac{3+\bar\gamma}{2+\bar\gamma}\underbrace{\sum_{k=0}^{+\infty}\frac{2^{-\left(k+1\right)}k!}{T_2\left(2\bar\gamma\right)\cdots\left[T_2\left(2\bar\gamma\right)+k\right]}}_{\Xi_2}\right)\\
&+\sum_{n=1}^{N_{\rm{r}}-1}A_{n,1}\sqrt{\frac{2\bar\gamma /(n+1)}{2+2\bar\gamma /(n+1)}}
\left(\underbrace{\sum_{k=0}^{+\infty}\frac{2^{-\left(k+1\right)}k!}{T_2\left(2\bar\gamma /(n+1)\right)\cdots\left[T_2\left(2\bar\gamma /(n+1)\right)+k\right]}}_{\Xi_3}\right).
\end{split}
\end{equation}
\hrulefill
\vspace*{-4pt}
\end{figure*}
Notably, $\Xi_1$ is acquired through the derivative of the series expansion in Equ. \eqref{eq26}, which is
\begin{equation}
\begin{split}
\frac{{\rm{d}}}{{\rm{d}}x}\left(\prod_{i=1}^{n}\left(\frac{1}{x+a_i}\right)\right)=\frac{-1}{\prod\limits_{i=1}^{n}{\left(x+a_i\right)}}\sum_{i=1}^{n}\left({\frac{1}{x+a_i}}\right).
\end{split}
\end{equation}
Moreover, $\Xi_2$ and $\Xi_3$ can be easily derived by expanding $\beta$-function.

\section{Simulation}
\label{section5}

In this part, numerical results will be provided to examine the feasibility and validity of the former derivations. And it should be noticed that all the following experiments are based on BPSK modulation. Besides, the ergodic mutual information in the simulation results is measured by ``bits/symbol''.

\begin{figure}[!t] 
\setlength{\abovecaptionskip}{-4pt} 
\centering 
\includegraphics[width=0.45\textwidth]{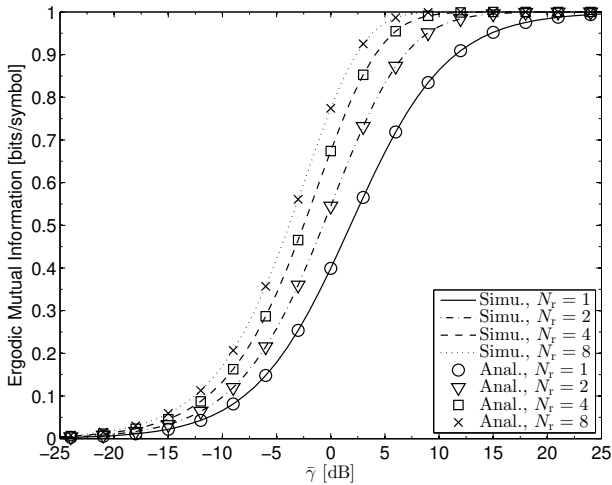} 
\caption{Simulated and analytical ergodic mutual information versus $\bar\gamma$ when $L=1$.}
\label{figure1}
\vspace{-4pt}
\end{figure}

\begin{figure}[!t] 
\setlength{\abovecaptionskip}{-4pt} 
\centering 
\includegraphics[width=0.45\textwidth]{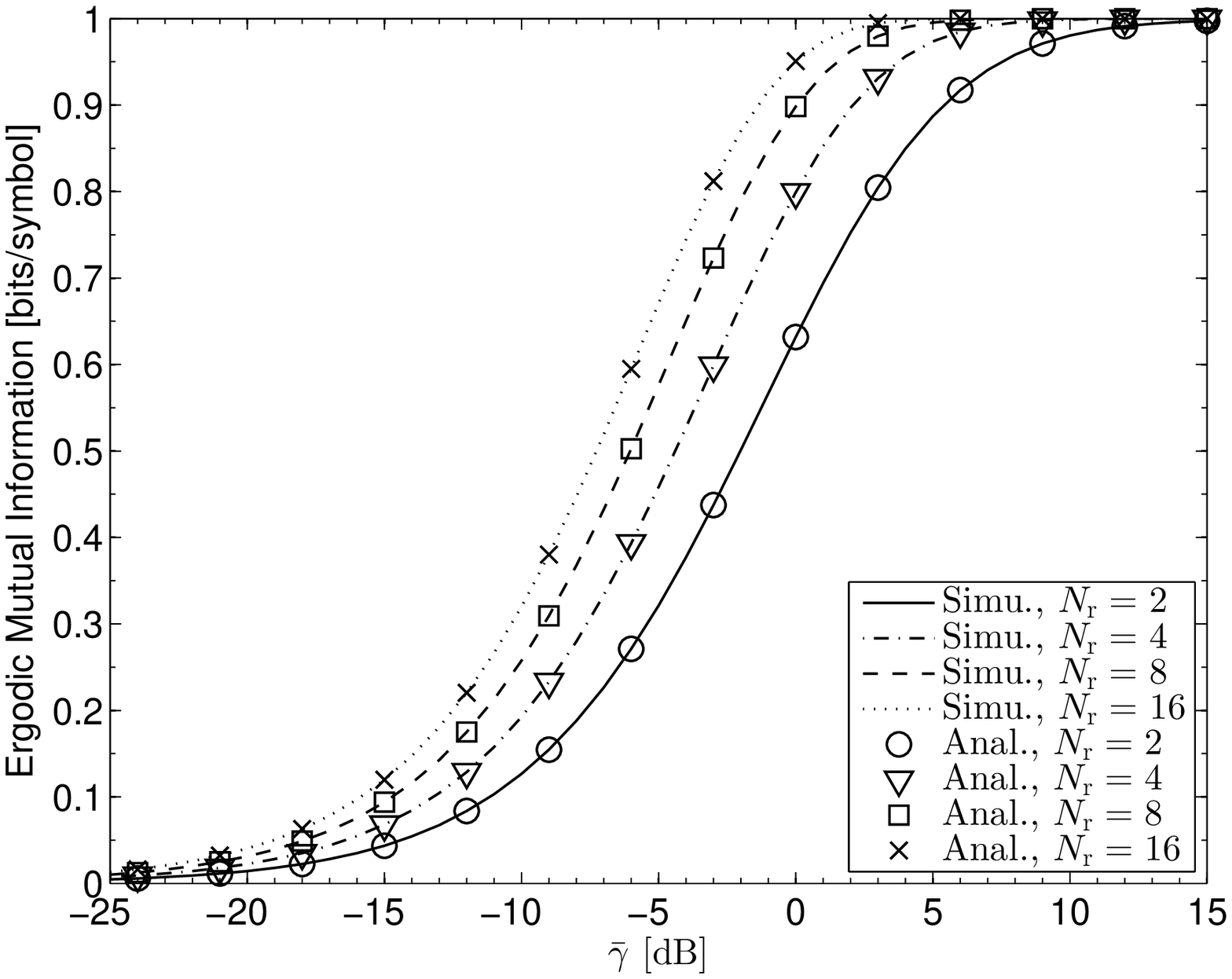} 
\caption{Simulated and analytical ergodic mutual information versus $\bar\gamma$ when $L=2$.}
\label{figure2}
\vspace{-4pt}
\end{figure}

Fig. \ref{figure1} compares the simulated and analytical ergodic mutual information in terms of $\bar\gamma$ for selected values of $N_{\rm{r}}$ when $L=1$. The simulated results are all obtained on the basis of Monte-Carlo experiments and the analytical values are calculated by Equ. \eqref{eq27}. As explained earlier, the expansion of the $\beta$-function in Equ. \eqref{eq27} converges much faster than the definition formula, thus it is enough to sum up the first $K$ terms in the expansion for highly accurate computation of the mutual information. In our simulations, $K$ is fixed to be 10. Most importantly, it can be observed  that the analytical results match well with the simulations for the curves and circles almost coincided with each other, which verifies the validity of the previous deduction. Another important observation is that the largest ergodic mutual information of BPSK is limited by 1 bits/symbol due to the discrete inputs, which is totally different from that of Gaussian inputs. 
Fig. \ref{figure2} provides the numerical results of $L=2$. The analytical mutual information is calculated by Equ. \eqref{eq33}. As can be seen from this graph, the analytical values also fit well with the simulated values, which again supports the former derivations.

\begin{figure}[!t]
\setlength{\abovecaptionskip}{-4pt}  
\centering 
\includegraphics[width=0.45\textwidth]{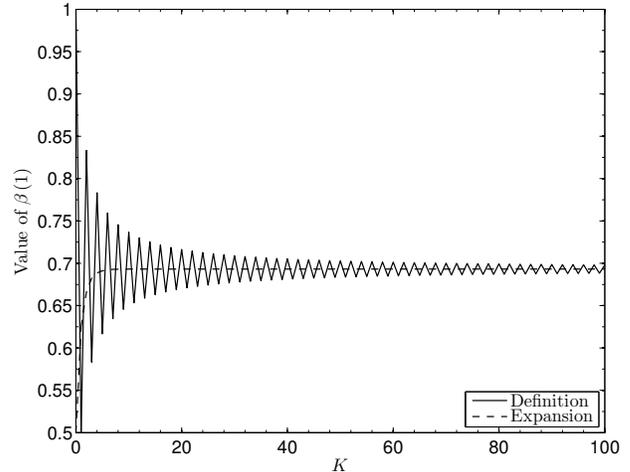} 
\caption{Convergence of the definition and expansion of $\beta\left(x\right)$ in Equ. \eqref{eq26} when $x=1$.}
\label{figure3}
\vspace{-4pt}
\end{figure}

\begin{figure}[!t] 
\setlength{\abovecaptionskip}{-4pt} 
\centering 
\includegraphics[width=0.45\textwidth]{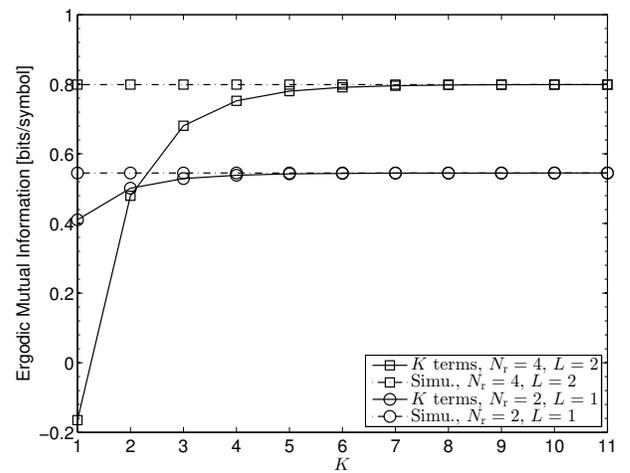} 
\caption{The change of ergodic mutual information with the increment of $K$ when $\bar\gamma=0$ dB and only.}
\label{figure4}
\vspace{-4pt}
\end{figure}

Next, let us turn to the convergence rate of the series expansion in Equ. \eqref{eq27} and Equ. \eqref{eq33}. As stated before, the expansion of $\beta$-function in Equ. \eqref{eq26} converges much faster than the definition formula. To further examine the convergence of the expansion, Fig. \ref{figure3} illustrates the change trend of the $\beta$-function with the number of summation terms, $K$, when $x=1$. As it shows, the expansion, denoted by dashed line, will converge to be constant when $K=6$. In contrast, the definition based result, denoted by solid lines, has not converged even though $K=100$. Taken together, it makes sense to calculate the mutual information with the expansion. Furthermore, Fig. \ref{figure4} plots the change of ergodic mutual information as a function of $K$ for different antenna deployment when $\bar\gamma=0$ dB. It is clear that the series expansions in Equ. \eqref{eq27} and Equ. \eqref{eq33} can approximate the real results well, thus it is accurate enough to utilize the summation of the first $K$ terms to estimate the exact values.        

\section{Conclusion}
\label{section6}
This paper investigates the ergoidc mutual information of H-S/MRC under BPSK/QPSK modulations and a general recursive formula is developed to analytically calculate the ergodic mutual information. Furthermore, based on this general formula, series expressions, with high precision, for the mutual information can be formulated once $L$ and $N_{\rm{r}}$ are fixed. Numerical results show that the series expansion has a fast-convergence rate and provides a
simple and numerically efficient way to calculate the ergodic mutual information of H-S/MRC systems.

\vspace{12pt}
\end{document}